\documentclass[aps,showpacs,tightenlines,twocolumn,nofootinbib,nobibnotes,superscriptaddress]{revtex4-1}
\makeatletter

\newcommand{\Rmnum}[1]{\expandafter\@slowromancap\romannumeral #1@}
\makeatother
\usepackage{graphicx, epstopdf, CJK}
\usepackage{dcolumn}
\usepackage{bm}
\usepackage[utf8]{inputenc}
\usepackage[T1]{fontenc}
\usepackage{booktabs, array, mathptmx, float, tabularx, booktabs, lipsum,bm, amsmath,multirow}
\usepackage{siunitx, xcolor}
\usepackage{color}
\usepackage{pifont}
\usepackage[colorlinks=true,linkcolor=blue,citecolor=blue, urlcolor=blue,bookmarks=false]{hyperref}
\usepackage{hyperref}
\usepackage{gensymb}
\hypersetup{colorlinks,urlcolor=blue,citecolor=blue,linkcolor=blue}
\usepackage[version=4]{mhchem}
\graphicspath{{figs/}{figsgaoerb/}}

\begin{document}	
	\title{Non-Hermitian quasicrystalline topological insulators}

	\author{Xiaolu Zhu}
	\affiliation{The Institute of Technological Sciences, Wuhan University, Wuhan 430072, China}

	\author{Tan Peng}
        \email{tanpeng@whu.edu.cn}
	\affiliation{Key Laboratory of Artificial Micro- and Nano-Structures of Ministry of Education, School of Physics and Technology, Wuhan University, Wuhan 430072, China}
	
\author{Fang Lyu}
	\affiliation{Key Laboratory of Artificial Micro- and Nano-Structures of Ministry of Education, School of Physics and Technology, Wuhan University, Wuhan 430072, China}

	\author{Wei Cao}
	\affiliation{The Institute of Technological Sciences, Wuhan University, Wuhan 430072, China}

        \author{Yue Hou}
	\affiliation{The Institute of Technological Sciences, Wuhan University, Wuhan 430072, China}

       \author{Rui Xiong}
        \email{xiongrui@whu.edu.cn}
	\affiliation{Key Laboratory of Artificial Micro- and Nano-Structures of Ministry of Education, School of Physics and Technology, Wuhan University, Wuhan 430072, China}

	\author{Ziyu Wang}
	\email{zywang@whu.edu.cn}
      \affiliation{The Institute of Technological Sciences, Wuhan University, Wuhan 430072, China}
	\affiliation{Key Laboratory of Artificial Micro- and Nano-Structures of Ministry of Education, School of Physics and Technology, Wuhan University, Wuhan 430072, China}
	
	\begin{abstract}
In recent years, the interplay between non-Hermiticity and band topology is expected to uncover numerous novel physical phenomena. However, the majority of research has focused on periodic crystalline structures, with comparatively fewer studies exploring quasicrystalline systems. In this paper, we delve into the influence of asymmetric hopping on the topological insulators, specifically focusing on quantum spin Hall insulators and higher-order topological insulators in an octagonal Ammann-Beenker quasicrystalline lattice. We demonstrate that asymmetric hopping can significantly alter the distribution of edge states, leading to a uniform distribution across all boundaries or localizing them at a single edge, depending on the symmetry adjustments. Furthermore, we explore the robustness of higher-order topological corner states under perturbations, showing that these states can maintain their distribution even in the presence of non-Hermiticity. Our findings not only expand the current understanding of topological states in quasicrystals under non-Hermitian conditions, but also provide valuable theoretical guidance for manipulating corner state distributions in non-Hermitian quasicrystalline higher-order topological insulators.		 			
	\end{abstract}
	
	
	\maketitle
	
	\section{Introduction}
The past few decades have witnessed an explosion of interest in the unique properties of topological insulators that feature fully formed bulk gaps while hosting robust boundary states protected by bulk topology \cite{1,2,3,4,5,6,7,8,9,10}. In recent years, researches on topological states have expanded from periodic crystalline systems to quasicrystalline systems \cite{11,12,13,14,15,16,A1,A2,A3}, which possess forbidden symmetries in crystals, such as the fivefold, eightfold, and twelvefold rotational symmetries. Some typical topological insulator phases including topological quantum spin Hall insulator \cite{15,16}, Chern insulator \cite{17,Han}, topological semimetal \cite{18,PhysRevB.108.195306}, topological superconductor \cite{19,PhysRevB.104.144511,PhysRevLett.116.257002} and topological Anderson insulator \cite{20,PhysRevB.103.L241106,PhysRevB.100.115311}, have been proposed in quasicrystals. More strikingly, higher-order topological insulators (HOTIs) protected by high-rotational symmetries have been discovered in quasicrystals \cite{11,A4,A5}. Thus, the unique properties of quasicrystals provide an ideal platform for exploring novel topological states.

The Hermiticity of Hamiltonian has always been a prerequisite for the use of quantum mechanics, which guarantees the conservation of probability of isolated quantum systems and the real value of the expected energy relative to the quantum state. However, it is common in nature for particles or energies to be exchanged between external degrees of freedom outside Hilbert space, so that probability is no longer conserved in such open systems \cite{A7}. And systems under non-Hermitian descriptions will have a complex energy spectrum, which are the embodiment of “the exchange”. Recently, due to the widespread presence of non-Hermiticity in natural and artificial materials, and the suitable physical description provided by non-Hermitian Hamiltonians for open systems, researches on topological states of matter have expanded to open systems such as open quantum systems \cite{21,22,23,24,25,26}, electronic systems with interactions \cite{27,28,29}, and classical systems with gain or loss \cite{30,31,32,33,34,35,36,37,38,39,40,41,42}, becoming a hotspot in condensed matter physics. Several theoretical and experimental studies have reported that non-Hermiticity introduces novel topological phenomena with no counterpart in Hermitian cases, including Weyl exceptional rings \cite{43,44,45,46}, bulk Fermi arcs \cite{47, 48}, half-integer topological charges \cite{49, 50}, and non-Hermitian skin effect \cite{51,52,53,54}.

On the other hand, the classical topological states in Hermitian systems can still exist under non-Hermitian consideration. For instance, Lee $et$ $al$. proposed a one-dimensional non-Hermitian topological insulator and proved that the bulk-boundary correspondence is modified in the presence of non-Hermiticity \cite{55}. Chen $et$ $al$. proposed a non-Hermitian Chern insulator and studied study the Hall conductance \cite{56}. Recently, HOTIs phases have been theoretically proposed in non-Hermitian systems \cite{21, 54, 57, 58} and experimentally realized in circuit systems \cite{59}. It is noting that the great majority of the previous works about topological insulators under non-Hermitian consideration were discussed in crystalline systems, while the investigation in quasicrystalline systems remains lacking.

In this paper, we study the topological states including quantum spin Hall insulator and HOTIs in the octagonal Ammann-Beenker (AB) tiling quasicrystalline lattice (QL) with non-Hermitian consideration. We find that the distribution of topological boundaries can be tuned by changing the strength of the non-Hermitian term and the symmetry of the system. First, under non-Hermitian consideration, the quasicrystalline system can have a uniform distribution of edge states on eight boundaries, which is the same as the Hermitian system. In addition, by changing the Pauli matrix to adjust the symmetry, the uniform distribution can be broken, and the edge states will be localized at only one edge of the structure. We further extend this phenomenon to HOTIs. The higher-order topological corner states will have local non-uniformity through the synergistic effect of non-Hermiticity and mass term after the Wilson mass term is introduced. Under the non-trivial case, we study the corner states and their distribution under the combined action of $\gamma$ and Wilson mass term $g$ with different Pauli matrix factors. Moreover, after introducing perturbation, the system has a good robustness and the distribution of corner states can exist stably. Our research has expanded the investigation into how asymmetric hopping modulate the distribution of first-order and higher-order edge states in quasicrystals, offering theoretical insights into the manipulation of corner state distributions in non-Hermitian quasicrystalline HOTIs.

The rest of this paper is organized as follows. We introduce a quantum spin Hall insulator model and a higher-order insulator model in a two-dimensional octagonal quasicrystalline lattice and give the details of the numerical methods in Sec.~\ref{Sec2}. Then, we provide the numerical results of studying the distribution of edge states of the two models and the robustness study under perturbation in Sec.~\ref{Sec3} and Sec.\ref{Sec4}. Finally, we summarize our conclusions in Sec.~\ref{Sec5}.

\section{Model and method}
\label{Sec2}
\begin{figure}[tp]
	\centering
	\includegraphics[width=9cm]{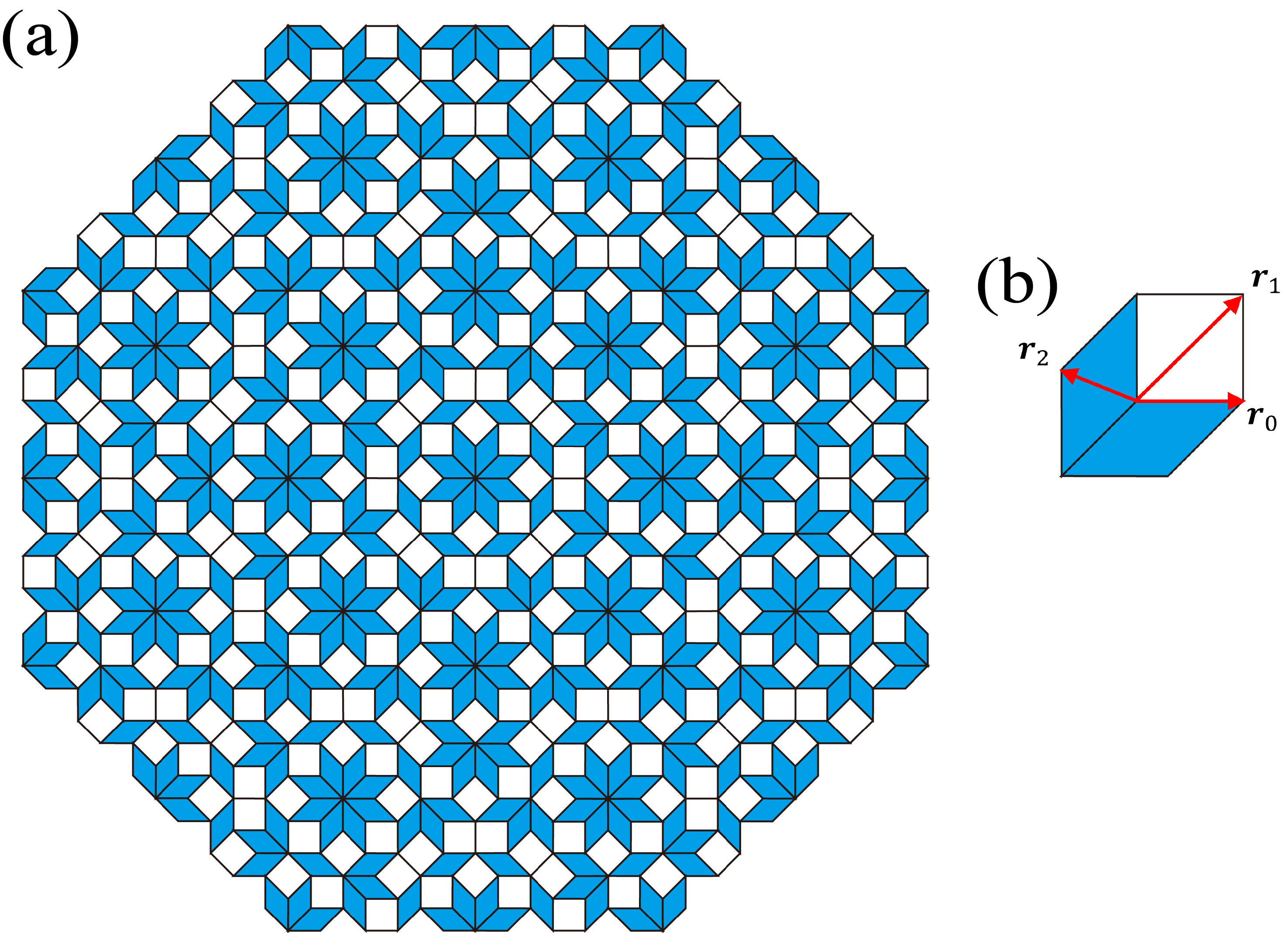}
	\caption{(a) The model of the AB tiling octagonal quasicrystal containing 833 vertices. The structure consists of a square and a diamond with a 45\degree angle. (b) Diagram of the first three nearest-neighbor hopping paths. The radio of the distances of the three bonds are $r_{0}:r_{1}:r_{2}=2\sin \frac{\pi }{8}:1:2\sin \frac{\pi }{4}$.}
	\label{fig:1}
\end{figure}

Starting from the non-Hermitian Bernevig-Hughes-Zhang (BHZ) model, we have constructed a non-Hermitian AB quasicrystal structure, as shown in Fig.~\ref{fig:1}(a). AB tiling QL is composed of square and rhombus with a 45\degree angle, containing 833 sites with eight-fold rotational symmetry. We consider the nearest-neighbor hopping, the next-nearest-neighbor hopping and the third-nearest-neighbor hopping, which correspond to the edge of the square, the short diagonal of the diamond and the diagonal of the square respectively, as shown in Fig.~\ref{fig:1}(b).

We define a non-Hermitian BHZ Hamiltonian on the lattice with the form of \cite{11}:

\begin{equation}\label{1}
\begin{aligned}
{H}_{0}=&\sum_{j\neq k}\hat{c}_{j}^{\dag}\{-\frac{f({r}_{jk})}{2}[i{t}_{1}({\sigma}_{3}{\tau}_{1}\cos{\psi}_{jk}+{\sigma}_{0}{\tau}_{2}\sin{\psi}_{jk})+{t}_{2}{\sigma}_{0}{\tau}_{3} \\
&-{\gamma}{\sigma}_{0}{\tau}_{3}\cos{\psi}_{jk}]\}\hat{c}_{k}+\sum_{j}\hat{c}_{j}^{\dag}[(M+2{t}_{2}){\sigma}_{0}{\tau}_{3}]\hat{c}_{j},
\end{aligned}
\end{equation}
where $\hat{c}_{j}^{\dag} = (\hat{c}_{j\alpha\uparrow}^{\dag},\hat{c}_{j\alpha\downarrow}^{\dag},\hat{c}_{j\beta\uparrow}^{\dag},\hat{c}_{j\beta\downarrow}^{\dag})$ represents the creation operator of an electron on the site j. $\alpha(\beta)$ is the index of orbitals, and $\uparrow(\downarrow)$ denotes the spin direction. The function $f({r}_{jk}) = {\Theta}({r}_{c}-{r}_{jk})\exp(1-{r}_{jk}/{\xi})$ represents the spatial decay factor of the hopping amplitude with $\xi$ as the decay length, ${r}_{jk} = \left|{r}_{j}-{r}_{k}\right|$. The factor ${\Theta}({r}_{c}-{r}_{jk})$ introduces a hard cutoff with hopping strength ${r}_{c}$. ${t}_{1}$ and ${t}_{2}$ hopping amplitudes. The ${\xi}_{1-3}$ and ${\tau}_{1-3}$ are the Pauli matrices acting on the spin and orbital degree of freedom, respectively. ${\sigma}_{0}$ and ${\tau}_{0}$ are 2${\times}$2 identity matrices. ${\psi}_{jk}$ describes the polar angle made by the bond between site \textit{j} and \textit{k} with respect to the horizontal direction \cite{60,61}. The term with ${\gamma}$ leads to the asymmetric hopping of the model, which is a description of the non-Hermitian strength of the system. When $\gamma = 0$, ${H}_{0}$ degenerates to a Hermitian Hamiltonian which describes a quantum spin Hall insulator phase in QL, respecting time-reversal, particle-hole and chiral symmetries \cite{60,61,66}. However, when $\gamma = 1$ , ${H}_{0}$ exhibit different symmetry classes compared with Hermitian systems \cite{PhysRevB.99.235112,PhysRevX.9.041015}. The model Hamiltonian preserves respects time-reversal symmetry $T = {i\sigma}_{2}{\tau}_{0}K$,variants of particle-hole symmetry $P' = {\sigma}_{3}{\tau}_{1}K$.
 \begin{figure}[tp]
	\centering
	\includegraphics[width=9cm]{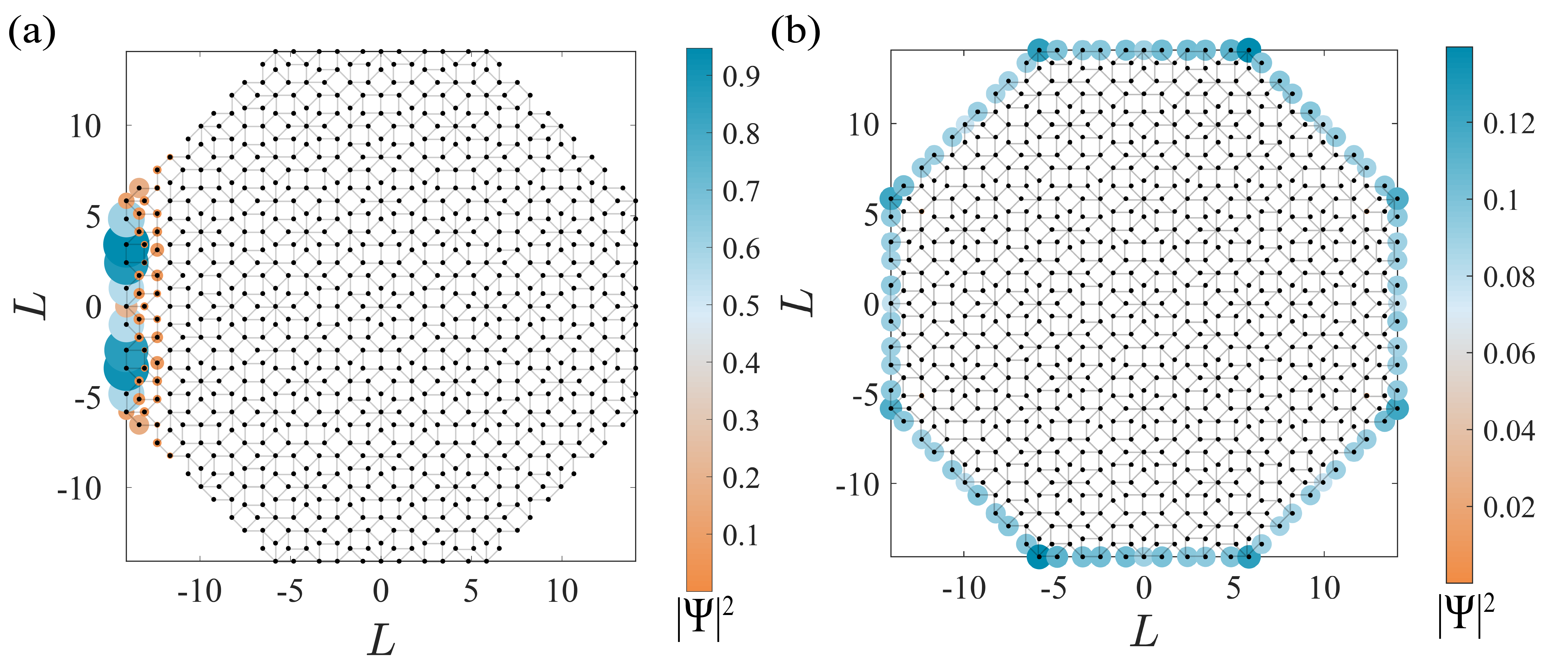}
	\caption{Wavefunction probability density $\sum\left|{\Psi}\right|^2$ when the non-Hermitian intensity (a) $\gamma$ = 1.5 with $\gamma{\sigma}_{0}{\tau}_{3}$, the edge state is localized on the left edge of the lattice; (b) $\gamma$ = 0.5 with $\gamma{\sigma}_{1}{\tau}_{1}$, the edge states are uniformly distributed on the eight edges.}
       \label{fig:2}
\end{figure}

\section{Edge states in quasicrystal}
\label{Sec3}
In this section, we focus on the topological edge or corner states in an AB tiling quasicrystal with the octagonal boundary conditions in response to asymmetric hopping. As shown in Fig.~\ref{fig:2}(a), the original edge state is localized on the left edge of the structure, which means that the traditional bulk-edge correspondence is broken. Nonetheless, our research indicates that by fine-tuning the Pauli matrix and the non-Hermitian intensity ($\gamma$ = 0.5, $\gamma{\sigma}_{1}{\tau}_{1}$), it is possible to achieve a uniform distribution of edge states across the eight edges of the quasicrystalline lattice, plotted in Fig.~\ref{fig:2}(b), which is consistent with the usual results in Hermitian cases. In this case, the systesm respects variants of particle-hole symmetry $P'={\sigma}_{3}{\tau}_{1}K$, variants of time-reversal symmetry $T'=i{\sigma}_{2}{\tau}_{0}T$, chiral symmetry $S=T'P'$, and pseudo-Hermiticity symmetry ${\sigma}_{0}{\tau}_{3}K$. We suppose that chiral symmetry is the main reason for the uniform distribution of edge states.

\begin{figure}[tp]
	\centering
	\includegraphics[width=9cm]{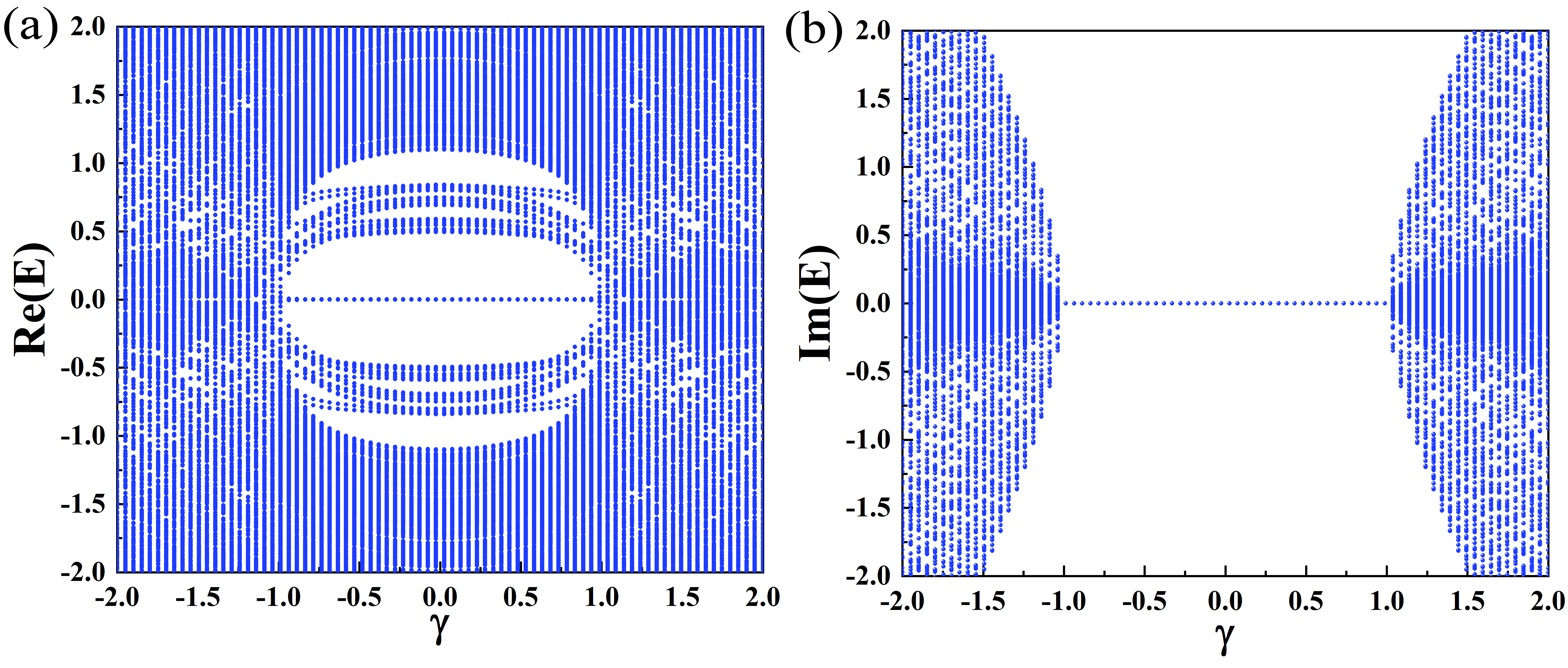}
	\caption{The energy spectrum of ${H}_{NH-BHZ}$ as a function of the non-Hermitian strength $\gamma$. (a) real part and (b) imaginary part of energy spectrums are plotted with ${M = -1,{t}_{1} = {t}_{2} = 1,{g} = 1}$. Due to the non-Hermitian action, the system behaves as a complex energy spectrum. When $\gamma$ takes the range from -1 to 1, the system is in a non-trivial state, and there are clear and stable topological ZEMs.}
	\label{fig:3}
\end{figure}

We also studied the impact of asymmetric hopping on higher-order corner states. The prerequisite is that the quantum spin Hall insulator on the QL can be converted into second-order topological insulator (SOTI) by introducing the Wilson mass term, which reads
\begin{equation}
{H}_{M}=g\sum_{j\neq k}\hat{c}_{j}^{\dag}[\frac{f({r}_{jk})}{2}{\sigma}_{2}{\tau}_{1}\cos(4{\psi}_{jk})]\hat{c}_{k},\label{2}
\end{equation}
where ${g}$ and $\cos(4{\psi}_{jk})$ represent the magnitude and the varying period of the Wilson mass, respectively. The mass term will break the time-reversal symmetry of the system which lead to cause the original helical edge states to transform into higher-order corner states, opening an energy gap. At the same time, the Pauli matrix corresponding to the non-Hermitian intensity in Eq (\ref{1}) is modified ($\gamma{\sigma}_{1}{\tau}_{1}\cos{\psi}_{jk}$) to achieve HOTI with eight corner states. Therefore, the Hamiltonian used to describe non-Hermitian SOTI should include Wilson mass as a whole, that is, ${H}_{NH-SOTI} = {H}_{0}+{H}_{M}$. The ${H}_{NH-SOTI}$ respects chiral symmetry $S=TP$. In order to obtain the effect of non-Hermitian strength $\gamma$ on the transformation of the system from trivial phase to non-trivial phase, we calculate the energy spectrum corresponding to different non-Hermitian strength, as shown in Fig.~\ref{fig:3}. We diagonalize ${H}_{NH-SOTI}$ under an open boundary condition (OBC) with ${M = -1, {t}_{1} = {t}_{2} = 1}$ and $\gamma$ takes the gradient from -2 to 2. In Fig.~\ref{fig:3}(a), it is found that ${\gamma}\approx1$ is the critical point at which topological phase transitions occur, and non-trivial zero-energy modes (ZEMs) appear. The critical point of the imaginary part occurs at the same locations as that of the real part, as shown in Fig.~\ref{fig:3}(b). In the topological non-trivial region, i,e. ${-1\lesssim\gamma\lesssim 1}$ the system hosts a pure real spectrum. In general, the mechanism to realize the real spectrum of Hermitian system is the protection of pseudo-Hermiticity symmetry (${\sigma}_{0}{\tau}_{3}K$). However, in our model, the mass term breaks the reciprocity and pseudo-Hermiticity that guarantee the system presents the pure real spectrum, which is also worth explotring.

\begin{figure*}[htbp!]
	\centering
	\includegraphics[width=18cm]{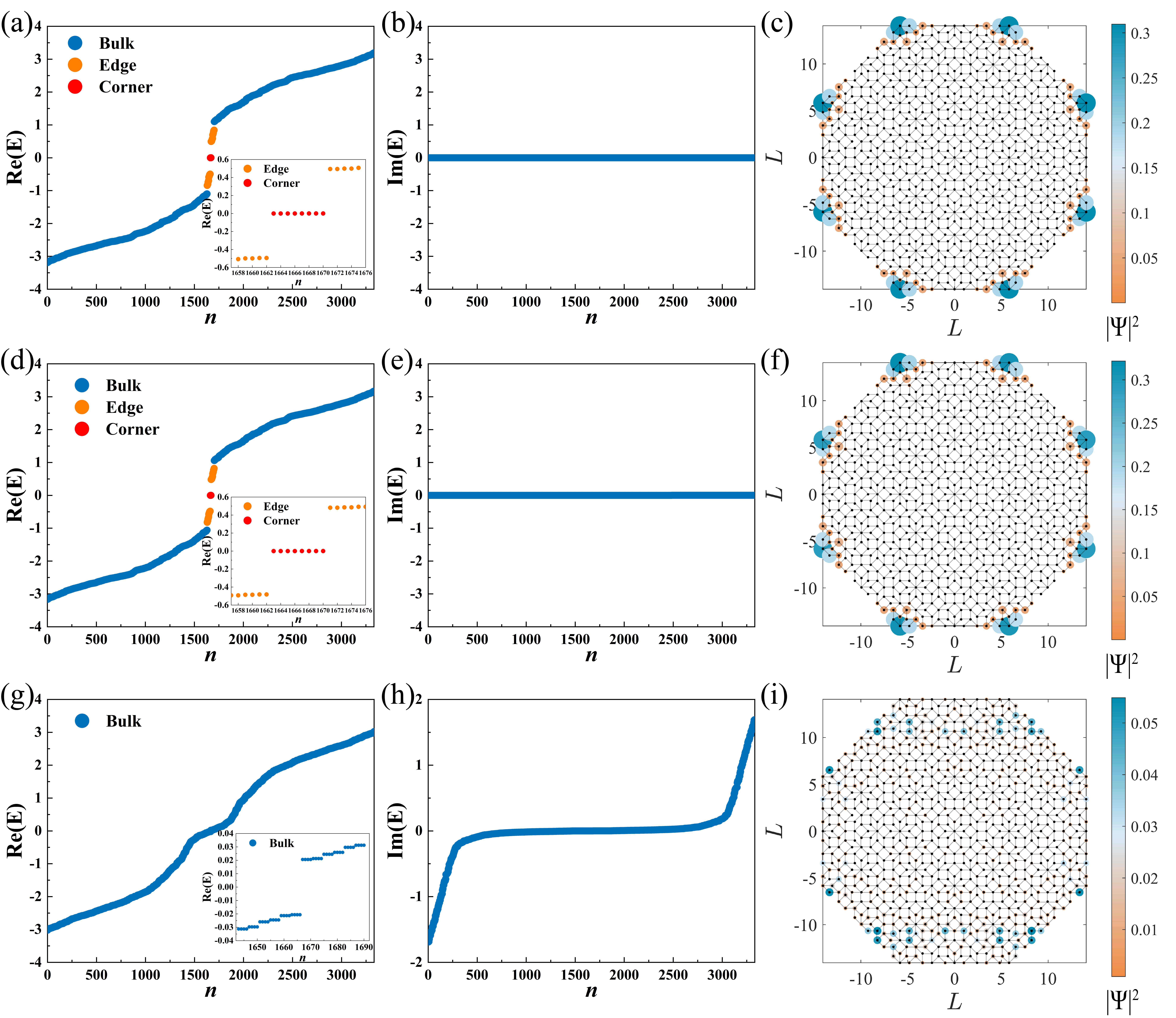}
	\caption{The complex energy spectrum and corner states of ${H}_{NH-SOTI}$ with ${M=-1,{t}_{1}={t}_{2}=1,{g}=1}$. (a), (d) and (g) are real part of energy spectrum with $\gamma=0,0.4,1.4$, respectively. And the insets in three figures show the enlarged section of eight ZEMs marked as the red dots. (b), (e) and (h) are imaginary part of energy spectrum with $\gamma=0,0.4,1.4$, respectively. When the system is in a non-trivial state, it behaves as a pure real spectrum until the topological phase transition occurs, which matches the rule in Fig.~\ref{fig:2}. (c), (f) and (i) are wavefunction probability density $\sum\left|{\Psi}\right|^2$. When $\gamma$ is in a certain range, the corner states of the system can exist stably. Once the topological phase transition occurs, the corner states are no longer maintained and spread into the bulk.}
	\label{fig:4}
\end{figure*}

In order to further understand the effect of non-Hermitian strength $\gamma$ on topological phase transitions, we contrast the energy spectrum including the real and imaginary part and the wavefunction probability for $\gamma = 0,0.4,1.4$, respectively, according to the information we get from Fig.~\ref{fig:3}. Without the intervention of non-Hermitian($\gamma = 0$), the energy spectrum shows that gapless in-gap states and edge states respectively marked by red and orange dots occupy the bulk gap under OBC in Fig.~\ref{fig:4}(a). The inset in Fig.~\ref{fig:4}(a) shows the enlarged section of eight ZEMs marked as the red dots. Because the system is under Hermitian condition($\gamma=0$), the energy spectrum of the imaginary part is zero, plotted in Fig.~\ref{fig:4}(b). It has been reported that eight ZEMs will appear in the system within the edge gap when the term Wilson mass is considered, that is, ${g}$ = 1 \cite{11}. The probability density of the in-gap eight eigenstates near zero energy shows that the eight corner states are located in the eight corners of the lattice, indicating the existence of SOTI phase, as verified from Fig.~\ref{fig:4}(c). When $\gamma=0.4$, ZEMs maintain stable, and the edge states and corner states coexist in the gap, as seen in Fig.~\ref{fig:4}(d). And inset of Fig.~\ref{fig:4}(d) shows the enlarged section of eight ZEMs marked as red dots. It is interesting to find that the value of the imaginary part of the energy spectrum is completely zero, as shown in Fig.~\ref{fig:4}(e). And the eight corner states appear on the eight angles respectively in Fig.~\ref{fig:4}(f). In combination with Fig.~\ref{fig:3}, the system will change from a topological state to a trivial state when $\gamma > 1$. Therefore, we set $\gamma = 1.4$ for calculation, plotted in Figs.~\ref{fig:4}(g), (h) and (i). In the Fig.~\ref{fig:4}(g), the ZEMs disappear, and a series of degenerate in-gap states emerge. The original corner state begins to spread into the bulk, and the non-trivial state of the system are no longer maintained, which can be seen in Fig.~\ref{fig:4}(i). At the same time, it can be seen in Fig.~\ref{fig:4}(h) that the phenomenon of the pure real part has disappeared. This result matches well with the evolution as function of $\gamma$ of the energy spectrum above. Further research will be carried out later on the premise that $\gamma = 0.4$ keeps the system in a non-Hermitian state.

In the absence of mass term, the system has edge states around the edges of the octagonal lattice. And the mass term is the cause of inducing the higher-order topological corner state of the system \cite{66}. The physical mechanism of topological ZEMs can be explained by Jackiw-Rebbi mechanism that topological ZEMs emerge when a mass domain wall forms \cite{67}. The Wilson mass term depends on the polar angle of the bond ${\psi}_{jk}$. However, the analytic expression of effective Wilson mass for edge states is difficult to get because of the lack of translation symmetry in quasicrystals. According to the reported research \cite{67}, the approximation that considering an edge of the boundary as a long “bond”, so that the sign of the effective Wilson mass depends on the orientation of the edge. It can be used as a rough decision rule to determine the positioning of corner states in polygonal quasicrystals. Observing the Eq({\ref{2}), it can be found that the non-Hermitian term and the Wilson mass term do not have the same matrix factors. The Pauli matrices they act on are ${\sigma}_{1}{\tau}_{1}$ and ${\sigma}_{2}{\tau}_{1}$, respectively. Therefore, we will define an adjusted Wilson mass term ${H}_{m}$:
\begin{equation}
{H}_{m}=g\sum_{j\neq k}\hat{c}_{j}^{\dag}[\frac{f({r}_{jk})}{2}{\sigma}_{1}{\tau}_{1}\cos(4{\psi}_{jk})]\hat{c}_{k}.\label{3}
\end{equation}

Under such circumstance, we will have the opportunity to explore the joint effects of $\gamma$ and $g$. In this case, the Hamiltonian of the system will be of the following form:
\begin{equation}
{H}_{NH-SOTI}^{*}={H}_{0}+{H}_{m}.\label{4}
\end{equation}

\begin{figure*}[htbp!]
	\centering
	\includegraphics[width=18cm]{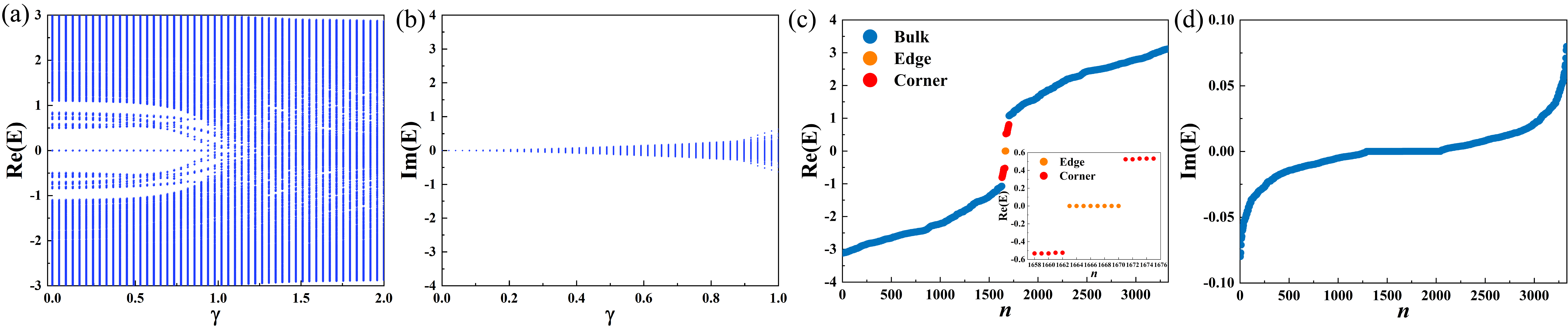}
	\caption{The complex energy spectrum of ${H}_{NH-SOTI}^{*}$ as a function of the non-Hermitian strength $\gamma$. (a) real part and (b) imaginary part of energy spectrums are plotted with ${M=-1, {t}_{1}={t}_{2}=1, g=1}$. Topological states are robust to $\gamma$ and in-gap ZEMs can exist stably. The topological phase transition occurs as $\gamma$ increases. The system presents complex spectrum characteristics due to the influence of non-Hermiticity, and can be regarded as pure real spectrum when $\gamma$ is weak. The energy spectrum obtaining real part (c) and imaginary part (d) of ${H}_{NH-SOTI}^{*}$ with ${M=-1, {t}_{1}={t}_{2}=1, g=1}$, $\gamma=0.4$. The inset of (c) shows the enlarged section of eight ZEMs marked as the red dots. The imaginary part of ${H}_{NH-SOTI}^{*}$ is non-zero.}
	\label{fig:5}
\end{figure*}

The Hamiltonian ${H}_{NH-SOTI}^{*}$ respects the particle-hole symmetry $P={\sigma}_{3}{\tau}_{1}T$. We calculate the energy spectrum of ${H}_{NH-SOTI}^{*}$ with different $\gamma$, and the real part of spectrum is shown in Fig.~\ref{fig:5}(a), it is found that with the increase of $\gamma$, the system will undergo topological phase transition at  $\gamma \approx 0.9$, father increase $\gamma$, the ZEMs disappear, indicating the system has become a pure insulator. With the increase of $\gamma$, the system will become a complex energy spectrum, although the real part is still in a topologically non-trivial state. It can be seen that the system is a pure real energy spectrum when $\gamma\lesssim0.2$, the imaginary part gradually becomes obvious with the increase of $\gamma$, as plotted in Fig.~\ref{fig:5}(b).

We diagonalize the ${H}_{NH-SOTI}^{*}$ under OBC with $\gamma$=0.4. The energy spectrums are plotted in Figs.~\ref{fig:5}(c) and (d). The inset of Fig.~\ref{fig:5}(c) shows that there are eight ZEMs with well stable in the real part of the energy spectrum. The posible resonalbe explanation that the value of energy spectrum is non-zero is that the pseudo-Hermiticity is broken \cite{68}, as shown in Fig.~\ref{fig:5}(d). The corner states, as depicted in Fig.~\ref{fig:6}, are located at four vertices, highlighting the uneven distribution of higher-order corner states. As expected, the interaction between $\gamma$ and $g$ has a significant effect on the topological corner states of the octagonal QL. This phenomenon is well described by the definition of hybrid skin-topological effects proposed by Li $et$ $al$., which describes the skin effect acting on the topological edge states \cite{PhysRevLett.128.223903,PhysRevLett.123.016805,PhysRevB.108.075122}. We will use the generalized Jackiw-Rebbi theorem to explain the corner modes. As shown in Fig.~\ref{fig:7}(b), the edges of QL are approximated by an octagon. The circular diagrams in Fig.~\ref{fig:7}(a) are used to show the symbol of the edge quality term, and the different color intervals correspond to the sign and values of $\cos(\eta{\psi}_{jk})$.

\begin{figure}[bp]
	\centering
	\includegraphics[width=8cm]{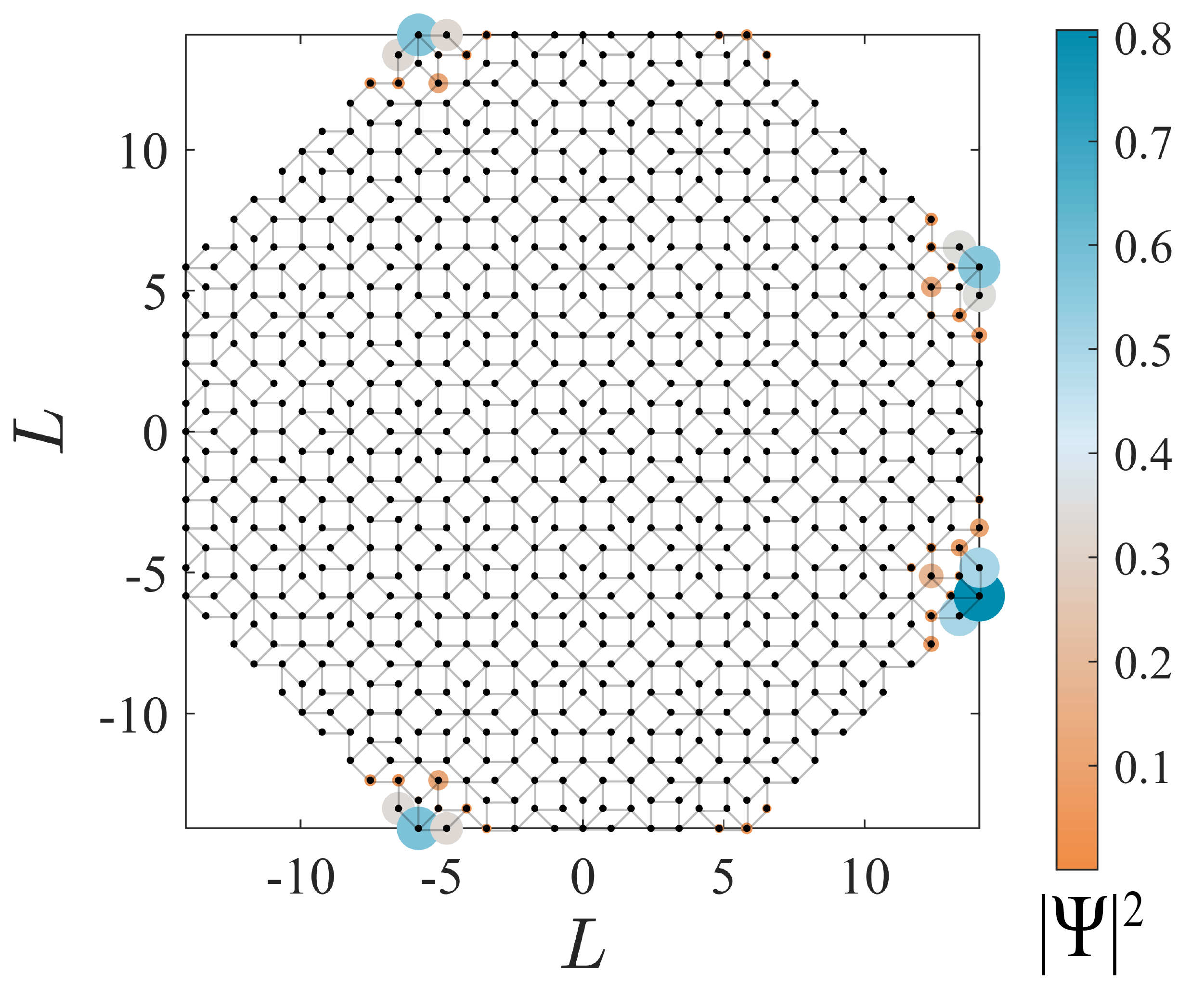}
	\caption{The wavefunction probability density $\sum\left|{\Psi}\right|^2$ of the ZEMs. Under the joint action of mass term and non-Hermiticity, the non-uniform distribution of corner states, locating at four vertices, appears in the system.}
	\label{fig:6}
\end{figure}

For example, the diagram above in Fig.~\ref{fig:7}(a) describes the factor $\cos(4{\psi}_{jk})$ in the mass term ${H}_{m}$. The mass term will be positive when the edge is in the red region $\psi\in(-\frac{\pi}{8},\frac{\pi}{8})\cup(\frac{3\pi}{8},\frac{5\pi}{8})\cup(\frac{7\pi}{8},\frac{9\pi}{8})\cup(\frac{11\pi}{8},\frac{13\pi}{8})$ and be negative when the edge is in the yellow region $\psi\in(\frac{\pi}{8},\frac{3\pi}{8})\cup(\frac{5\pi}{8},\frac{7\pi}{8})\cup(\frac{9\pi}{8},\frac{11\pi}{8})\cup(\frac{13\pi}{8},\frac{15\pi}{8})$. The other diagram in Fig.~\ref{fig:7}(a) describing the factor $\gamma\cos{\psi}_{jk}$, has the same definition: in the deep blue region which determined by $\psi\in(-\frac{\pi}{2},\frac{\pi}{2})$, the term including $\gamma$ is positive. And it will be negative in the wathet region is determined by $\psi\in(\frac{\pi}{2},\frac{3\pi}{2})$. The two edge of each corner forming octagonal QL are situated in different regions resulting in varying intensities of position reaching the corner through the two boundaries. For simplicity, we label the eight corners in Fig.~\ref{fig:7}(b) in clockwise order, denoting them I, II, III, …, VII, VIII. For example, the corner-VIII will be formed by electrons flowing towards the left at the top horizontal edge and moving up along the left oblique upward edge. The right edge of the corner-VIII forms an angle $\psi = 0$ from the horizontal direction, while the left edge of the corner-I forms an edge $\psi = \pi$ with the horizontal direction. The effective edge mass: $g\cos(4{\psi}_{jk}+\gamma\cos{\psi}_{jk})$ can be used to obtain the hopping strength to the vertices. Therefore, for the topmost boundary, the strength to the left is g+$\gamma (\psi = 0)$ and the strength to the right is $g-\gamma (\psi = \pi)$, indicating that the absolute magnitude is stronger to the left. For the upper left edge analysis, the two ends are corner-VII and corner-VIII. The left edge of corner-VIII forms an angle $\psi = \frac{3\pi}{2}$ from the horizontal direction. Hence, the strength of this edge toward corner-VIII is $-g-\frac{\sqrt{2}}{2}\gamma$ ($\psi = \frac{3\pi}{2}$). Accordingly, the strength of this edge towards corner-VII is $-g+\frac{\sqrt{2}}{2}\gamma$, because at corner-VII, the edge forms an angle $\psi=\frac{\pi}{4}$ to the horizontal direction. As a result, by analyzing the absolute magnitude of the hopping strength from two edges at each corner, it is possible to distinguish the corner with the more prominent aggregation intensity. As marked in Fig.~\ref{fig:7}(b), the effective intensity is stronger at the four vertices (II, III, V, VIII), and the probability density of these corner state should be enhanced, while the probability density of the other corners is suppressed. The matching results are shown in Fig.~\ref{fig:6}.
\begin{figure}[t]
	\centering
	\includegraphics[width=8cm]{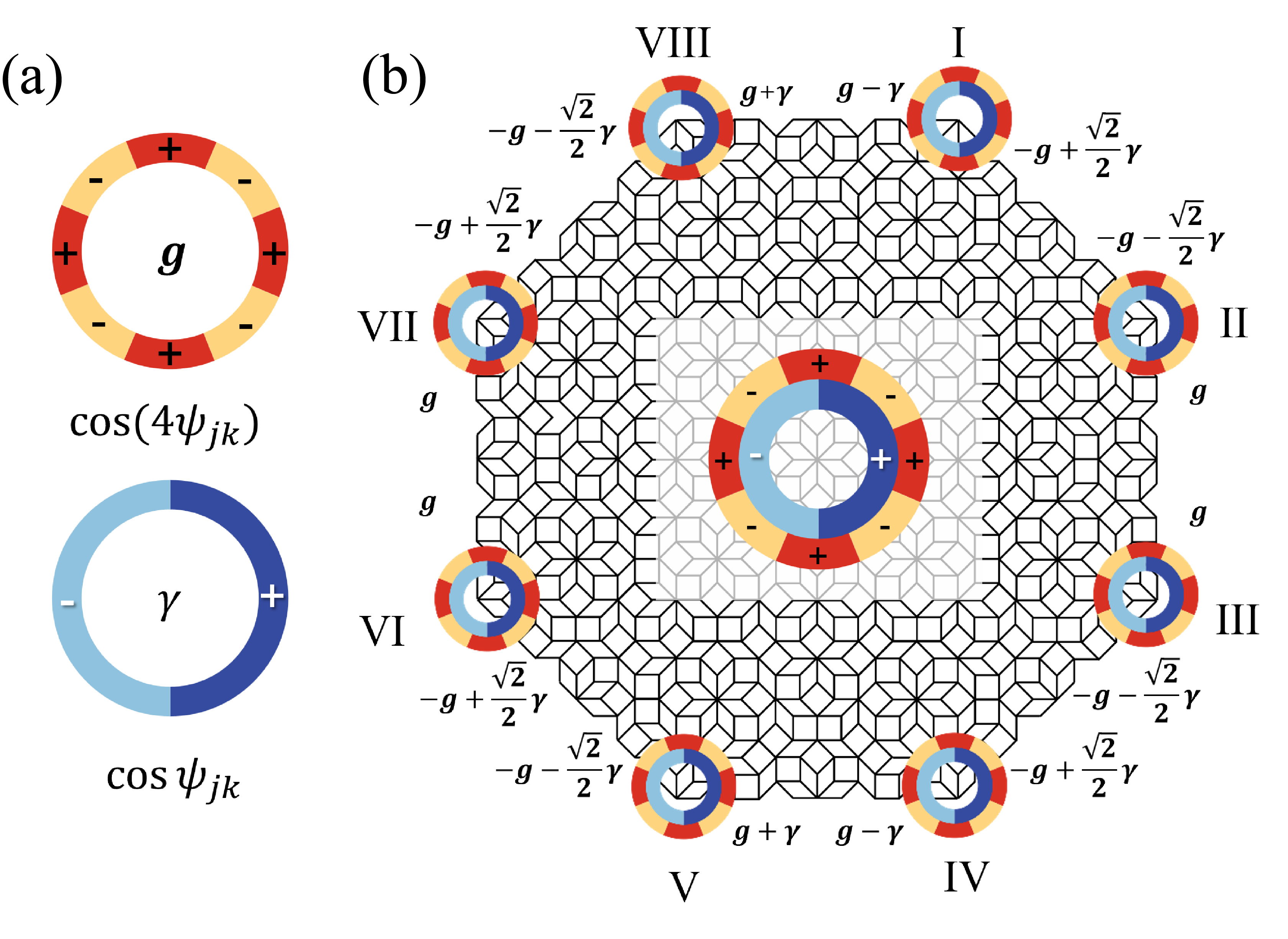}
	\caption{Scheme using the generalized Jackiw-Rebbi index theorem to account for the corner modes. (a) the up-circular chart depicting the sign of the mass term,$\cos{4\psi}_{jk}$. Red and yellow region denote positive and negative signs, respectively. The down-circular chart depicting the sign of the $\cos{\psi}_{jk}$ of $\gamma$, deep blue and wathet region denote positive and negative signs, respectively. (b) At each corner of the octagon two circles are shown nested together. The outside is used to calculate $g\cos{4\psi}_{jk}$ and the inside is used to calculate $\gamma\cos{\psi}_{jk}$. Corner will be increased if it is located in red and deep blue regions and decreased if it is located in yellow and wathet regions, resulting in different aggregation intensities.}
	\label{fig:7}
\end{figure}

To show that local corner states appear at different positions through modulation by considering different non-Hermitian hopping terms. We define the non-Hermitian hopping terms (NHHT) in Eq.(\ref{1}) as follows:
\begin{equation}
{H}_{NHHT}=\sum_{j\neq k}\hat{c}_{j}^{\dag}[\frac{f({r}_{jk})}{2}\gamma{\sigma}_{1}{\tau}_{1}\cos({\psi}_{jk})]\hat{c}_{k}.\label{5}
\end{equation}

Here, we set ${H}_{NHHT}$ as ${H}_{NHHT}^{*}$:
\begin{equation}
{H}_{NHHT}^*=\sum_{j\neq k}\hat{c}_{j}^{\dag}[\frac{f({r}_{jk})}{2}\gamma{\sigma}_{1}{\tau}_{1}\sin({\psi}_{jk})]\hat{c}_{k}.\label{6}
\end{equation}

Once more, we utilize the Jackiw-Rebbi mechanism to approximate the non-reciprocal hopping amplitude for each edge, as illustrated in Fig.~\ref{fig:8}(a). This approach induces edge states to four specific positions, underscoring the adjustable nature of the non-uniform corner state distribution in higher-order topological systems. The inset in Fig.~\ref{fig:8}(a) describes the symbols of $\cos{\psi}_{jk}$ and $\sin{\psi}_{jk}$ and their corresponding range of polar angles, which will produce a new nonreciprocal hopping effect under the new combination. In this case, the local intensity of four vertices (I, III, VI, VIII) are increased, and the other four vertices (II, IV, V, VII) are suppressed, as shown in Fig.~\ref{fig:8}(b). This finding underscores that a topologically non-trivial system can be engineered by tuning the non-Hermitian strength. Subsequently, by fine-tuning the matrix elements associated with the non-Hermitian strength and the Wilson mass term, one can modulate the local position of  corner states  to align with the desired location.

\begin{figure}[t]
	\centering
	\includegraphics[width=9cm]{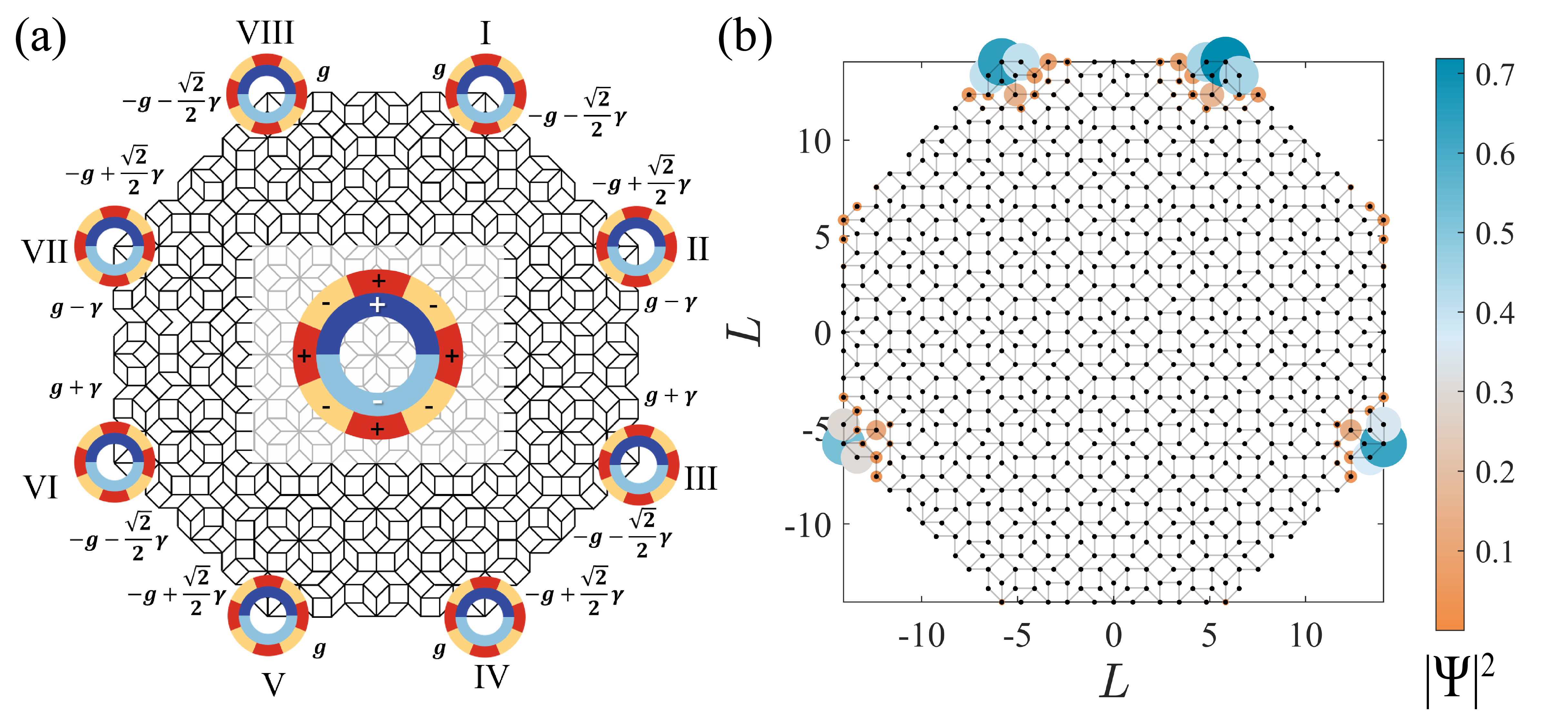}
	\caption{Scheme using the generalized Jackiw-Rebbi theorem to account for the corner modes.  (a) At each corner of the octagon, two circles are shown nested together. The outside is used to calculate $g\cos{4\psi}_{jk}$ and the inside is used to calculate $\gamma\sin{\psi}_{jk}$. The inset of (a) shows that red and yellow region denote positive and negative signs, deep blue and wathet region denote positive and negative signs, respectively. (b) The non-uniform distribution induced by ${H}_{NHHT}^*$. The positions of the corners of the eight ZEMS of the system are changed and localized in the new four vertices because of the changed phase factor $\sin{\psi}_{jk}$ of ${H}_{NHHT}^*$.}
	\label{fig:8}
\end{figure}

\section{Robustness study- Under perturbation}
\label{Sec4}
To assess the local stability of higher-order topological states post non-uniform distribution, we introduce a perturbation $w$ to examine the robustness of the system. The expression for the perturbation is as follows:
\begin{equation}
{H}_{w}=\sum_{j}\hat{c}_{j}^{\dag}(w{\sigma}_{0}{\tau}_{3})\hat{c}_{j}.\label{7}
\end{equation}
\begin{figure*}[htbp!]
	\centering
	\includegraphics[width=18cm]{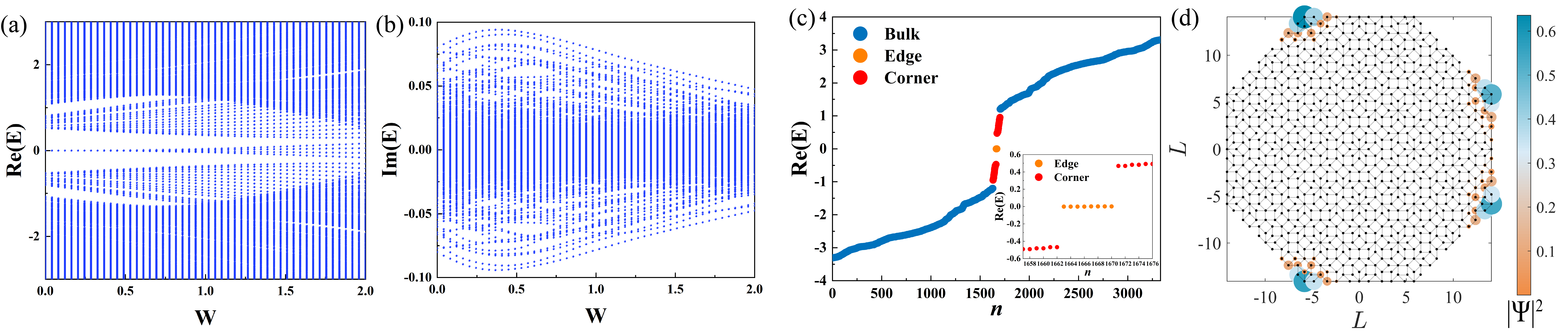}
	\caption{The complex energy spectrum containing (a) real and (b) imaginary part of the system as a function of the perturbation $w$. (c) When $w=0.3$, the system has eight ZEMs at OBC, and the inset of (c) shows the enlarged section of eight ZEMs marked as the red dots. (d) The wavefunction probability density $\sum\left|{\Psi}\right|^2$ of eight ZEMs. The non-uniform distribution of corner states has the same presentation under the weak perturbation.}
	\label{fig:9}
\end{figure*}

By calculating the energy spectrum under the perturbation strength $w$ gradient from 0 to 2, we obtain an interval range of the stability of the system, plotted in Fig.~\ref{fig:9}. In Fig.~\ref{fig:9}(a), it is found that when $w\lesssim0.5$, the ZEMs in the real part of the energy spectrum gap remain stable. As the perturbation strength gradually increases, the ZEMs in the gap begins to disappear and gradually spreads into the edge and bulk. Moreover, the value of imaginary energy spectrum of the system is non-Zero, shown in Fig.~\ref{fig:9}(b). We select $w = 0.3$ for detailed research, as shown in Fig.~\ref{fig:9}(c), eight ZEMs can be observed in the energy spectrum under open boundary conditions. It is a strong evidence of the robustness of topological corner states. We map out the wavefunction probability of eight ZEMs in Fig.~\ref{fig:9}(d). It is found that the original non-uniform distribution of corner states does not change, which is robust against perturbation. The robustness of the system under complex perturbation is further discussed in Appendix.

\section{CONCLUSION}
\label{Sec5}
In this work, we study the topological states including quantum spin Hall insulator and HOTIs in the octagonal AB tiling QL under non-Hermitian consideration. We find that the distribution of topological edge states can be tuned by changing the intensity of the non-Hermitian terms and the symmetry of the system. First, under the non-Hermitian consideration, the quasicrystalline system can have a uniform distribution of edge states on eight boundaries. Additionally, by altering the Pauli matrices to adjust symmetry, the  edge states localized at only one edge of the structure. Based on symmetry analysis, we think that chiral symmetry is the main factor affecting whether the eight edge states are uniformly distributed. We further extend this phenomenon to HOTIs. It can be found that the topological phase transition can be realized with the increase of non-Hermitian strength and the topological corner states gradually diffuse into the bulk. We find that the eight corner states are localized at the four vertices under the action of different hopping intensities on the eight boundaries. Meanwhile, by matching the phase factors of different mass terms, corner states can be obtained at different vertices, which provides theoretical guidance for realizing the local position control of the corner states of non-Hermitian systems and making them locate at the expected position. The uniformly distribution of topological edge (corner) states can be explained by the hybrid skin-topological effect, where the skin effect specifically acts only on the boundary states.

Moreover, after introducing perturbation, the setting of the perturbation term ensures that the symmetry protecting the system's topological is not broken, while also maintaining the system within the range of topological phase parameters by controlling the intensity of the perturbation term. The non-uniform distribution of corner states can exist stably. However, the on-site non-Hermitian term representing gain from the perturbation will interfere with the distribution of higher-order topological states of the system and cannot be maintained properly. 

Recently, the hybrid higher-order skin-topological effect has been successfully implemented in a specially designed network of non-reciprocal topological circuits \cite{000729179400003}. Moreover, the quasicrystalline quadrupole topological insulators has been experimentally realized in electrical circuits \cite{000659020500001}. Therefore, we believe that the non-Hermitian quasicrystalline topological insulators can be realized in the non-reciprocal topolectrical circuits.

\section{ACKNOWLEDGMENTS}
This work was supported by a project from the NSFC(Grant No. 12122408) and the National Key R\&D Program (Grant No. 2023YFB4603800). T.P. was supported by the Doctoral Research Start-Up Fund of Hubei University of Automotive Technology (Grant No. BK202216) and the Research Foundation of Hubei Educational Committee (Grant No. Q20231807).

\section{APPENDIX: ENERGY SPECTRUM AND WAVE FUNCTION UNDER THE COMPLEX PERTURBATION}
We try to introduce an on-site non-Hermitian term representing gain from the perturbation term, so we calculate the energy spectrum and wavefunction probability under the complex perturbation $iw$. It is described as follows:

\begin{equation}
{H}_{w}^{*}=\sum_{j}\hat{c}_{j}^{\dag}(iw{\sigma}_{0}{\tau}_{3})\hat{c}_{j}\label{8}.
\end{equation}

\begin{figure*}[htbp!]
	\centering
	\includegraphics[width=18cm]{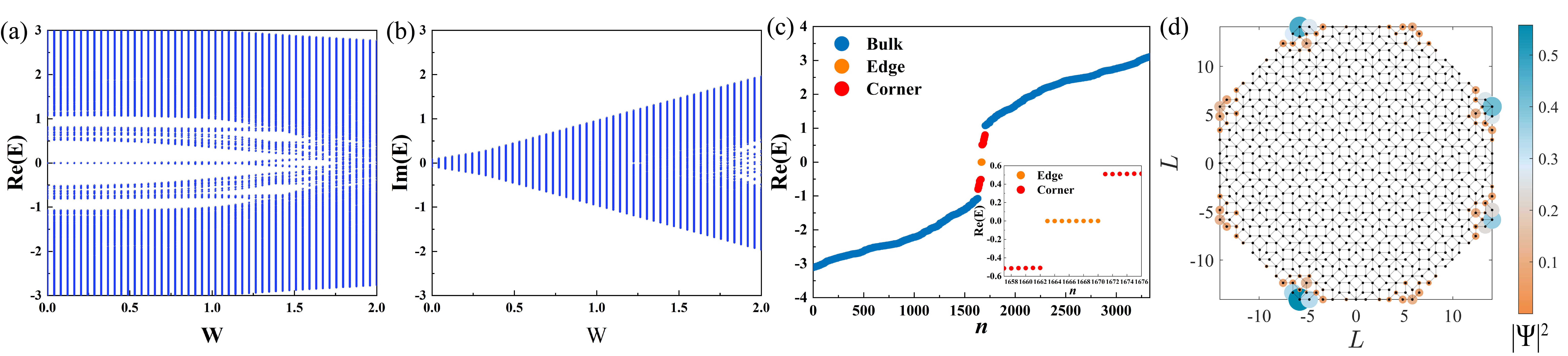}
	\caption{The complex energy spectrum containing (a) real and (b) imaginary part of the system as a function of the perturbation $w$. The term of perturbation is set as a complex. (c) When $w=0.3i$, the system has eight ZEMs at OBC, and the inset of (c) shows the enlarged section of eight ZEMs marked as the red dots. (d) The wavefunction probability density $\sum\left|{\Psi}\right|^2$ of eight ZEMs.}
	\label{fig:10}
\end{figure*}

Figures.~\ref{fig:10}(a) and (b) describe the real and imaginary part of the energy spectrum with perturbation $iw$, respectively. It is found that the system can exist topological non-trivial state under weak complex perturbations, and it always shows the property of complex energy spectrum. As shown in Fig.~\ref{fig:10}(c), there are still in-gap eight ZEMs which is the strong evidence of the robustness. We calculate the wavefunction probability of the eight modes closest to zero energy, as plotted in Fig.~\ref{fig:10}(d). It is found that the main distribution position of the ZEMs is still the four vertices (II, III, V, VIII) analyzed by Jackiw-Rebbi mechanism, but there is also weak intensity distribution in the remaining corners, which may be caused by the interaction between the original non-Hermitian term and complex perturbation.
	

%

\end{document}